\documentclass[
reprint,
amsmath,amssymb,
aps,
superscriptaddress,
]{revtex4-2}

\usepackage{graphicx}
\usepackage{dcolumn}
\usepackage{bm}
\usepackage{commath}
\usepackage{amsmath}
\usepackage{textcomp}
\usepackage{graphicx}
\usepackage[]{wasysym}
\usepackage{ulem}

\usepackage{hyperref}       
\usepackage{url}            
\usepackage{booktabs}       
\usepackage{amsfonts}       
\usepackage{nicefrac}       
\usepackage{microtype}      
\usepackage{lipsum}
\usepackage{color}
\renewcommand{\figurename}{Fig.}

\newcommand{\cea}{\affiliation{Universit\'e Paris-Saclay, CEA, CNRS, LIDYL, 91191, Gif-sur-Yvette, France}}

\newcommand{\queen}{\affiliation{School of Information Technology and Electrical Engineering, The University of Queensland, Brisbane, QLD, 4072, Australia}}

\newcommand{\caltech}{\affiliation{Department of Electrical Engineering, California Institute of Technology, Pasadena, California 91125, USA}} 

\newcommand{\hba}{\affiliation{Laboratoire Kastler Brossel, ENS-Universit\'e PSL, CNRS, Sorbonne Université, Coll\'ege de France, 24 rue Lhomond, 75005 Paris, France}}

\begin{document}

\preprint{APS/123-QED}

\author{Antoine Boniface}
\email{antoine.boniface@lkb.ens.fr}
\hba
\author{Mickael Mounaix}
\hba \queen
\author{Baptiste Blochet}
\hba \caltech
\author{Hilton B. de Aguiar}
\hba
\author{Fabien Qu\'er\'e}
\cea
\author{Sylvain Gigan}
\hba

\title{Spectrally-resolved point-spread-function engineering using a complex medium}

\begin{abstract}
Propagation of an ultrashort pulse of light through strongly scattering media generates an intricate spatio-spectral speckle that can be described by means of the multi-spectral transmission matrix (MSTM). In conjunction with a spatial light modulator, the MSTM enables the manipulation of the pulse leaving the medium; in particular focusing it at any desired spatial position and/or time. Here, we demonstrate how to engineer the point-spread-function of the focused beam both spatially and spectrally, from the measured MSTM. It consists in numerically filtering the spatial content at each wavelength of the matrix prior to focusing. We experimentally report on the versatility of the technique through several examples, in particular as an alternative to simultaneous spatial and temporal focusing, with potential applications in multiphoton microscopy. 
\end{abstract}
\maketitle
Temporal control of ultrashort pulses is a cornerstone of ultrafast optics. Pulse shaping refers to technologies that enable programmable reshaping of ultrafast optical waveforms, with control of phase, amplitude, and polarization \cite{weiner2000femtosecond}. It is now widely used in laser control over molecular and material responses \cite{weiner1990femtosecond}, but also for wavelength selective switches \cite{baxter2006highly} or in multiphoton microscopy to adjust the image contrast or resolution \cite{silberberg2009quantum}. In general, the method relies on spatially dispersing a pulse with a diffraction grating in order to manipulate separately its frequency components. 

Recent advances in very different research fields have shown that the ability to manipulate the full spatio-spectral or spatio-temporal structure of laser pulses, i.e. to introduce spatio-spectral couplings, can open new possibilities to control light propagation \cite{sun2018four} and light-matter interaction \cite{PhysRevLett.108.113904,pariente2015spatio,sainte2017controlling, PhysRevLett.124.134802}. Another important example of the usefulness of spatio-temporal couplings is given by simultaneous space-time focusing \cite{Zhu:05, Oron:05}, now widely used in multiphoton excitation for neuroscience \cite{papagiakoumou2020scanless}. Although some extensions of temporal shapers to spatio-temporal control have been demonstrated \cite{Feurer:02, Vaughan:02,feurer2003spatiotemporal}, such schemes are optically very complex, and their use has thus not become widespread. A general-purpose versatile method for spatio-spectral pulse control is still elusive.

Scattering of broadband light in disordered materials randomly mixes the spatial and spectral modes of the incident pulse. When a coherent ultrashort pulse travels through a multiply scattering medium its optical wavefront is spatially distorted and forms a speckle pattern \cite{goodman2007speckle}. When the bandwidth of the laser $\Delta\lambda^{laser}$ is broader than the spectral correlation bandwidth of the medium $\delta\lambda_m$, the speckle depends on the wavelength $\lambda$. Therefore, the scattering medium acts as a dispersive optical element for ultrashort pulses of light. In this regime, the optical transformation of the field induced by the medium is very complex but still remains linear and deterministic, hence controllable. Owing to the availability of spatial light modulators (SLMs), several techniques based on wavefront shaping were developed to experimentally characterize this process. A recurrent application is to find the incident wavefront that counterbalance the effects of scattering and thus re-compress the pulse to its initial duration and focus it on a diffraction limited spot. For instance it can be achieved by iteratively optimizing the incident wavefront \cite{PhysRevLett.106.103901,katz2011focusing} but also by using digital phase conjugation \cite{morales2015delivery}, spectral pulse shaping \cite{mccabe2011spatio}, or time-gating techniques \cite{Mounaix2016a}. Another and more global approach to describe and manipulate the outgoing broadband light consists of measuring the multi-spectral transmission matrix (MSTM) \cite{Andreoli2015}. The MSTM is a set of $N_\lambda = \Delta\lambda^{laser} / \delta\lambda_m$ monochromatic transmission matrices (TMs); each TM linearly relates the input field to the output field of the medium \cite{popoff2010measuring} for a given spectral component of the pulse. 
The full set of matrices provides both spatial and spectral/temporal control of the transmitted pulse; in particular enhancing a single spectral component of the output pulse or focusing it at a given time can be performed \cite{Mounaix2016,Mounaix:18,boniface2019rapid}.

The key point here is that these techniques manipulate both spatial and spectral degrees of freedom of the pulse by only using a single SLM. This is possible thanks to the spatio-spectral coupling resulting from the propagation through the medium. This is what we will exploit here, now to implement 3D spatio-spectral control, relying on a single SLM.
Although pulse control in complex media has already been studied in the last years, to our knowledge, the spatio-temporal degrees of freedom of a scattering medium have never been used to \textit{spectrally} engineer the point-spread function (PSF) of an ultrashort pulse.

Here, we exploit the MSTM in conjunction with a single SLM to perform arbitrary spatio-spectral PSF engineering. It consists in \textit{(i)} measuring the MSTM to characterize light scattering induced by the medium and \textit{(ii)} numerically filtering a virtual pupil function with a spectrally-dependent mask prior to \textit{(iii)} focusing. Importantly, nearly any arbitrary mask (phase and/or amplitude) can be applied onto the pupil function. This versatility is experimentally shown through the generation of two different spatio-spectral PSFs that both aim at decoupling axial confinement from lateral extent, with multiphoton microscopy applications envisioned. First, we revisit traditional temporal focusing (TF) and show that our approach is not restricted to disperse the pulse along only one spatial dimension as offered by diffraction gratings. Secondly, we report on another TF-like PSF that benefits from the high transverse resolution of a bessel beam but with a better axial confinement. Corresponding spatio-temporal profiles are characterized with a 2-photon fluorescence process.
\newline

Focusing light spatially implies that all incident contributions, or k-vectors, arrive at the focal point with the same phase. Stated differently, the Fourier transform of the field in the focal plane -- henceforth referred to as the pupil function -- has a flat phase.
Any modification of this pupil function (in phase and/or amplitude) impacts on the PSF of the optical system. Without a scattering medium, the pupil function can be engineered by displaying a phase mask onto a SLM placed in a plane conjugate to the back aperture of the illumination objective. In the presence of a scattering medium, this operation is pointless since all the information would be scrambled before reaching the focal plane. 
In this regime, we need to combine wavefront shaping and PSF engineering. 

\begin{figure}[t!]
\centering
\includegraphics[width=\linewidth]{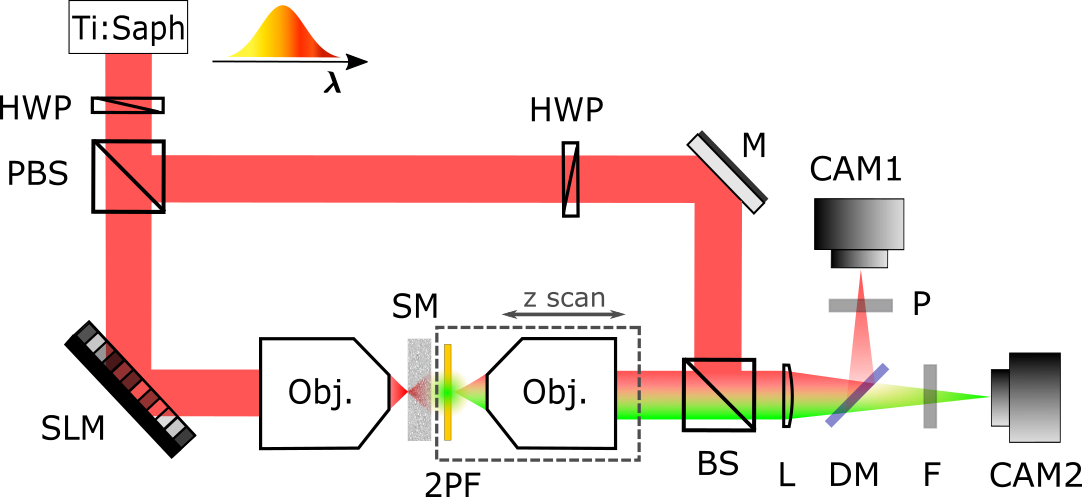}
\caption{Experimental setup scheme. HWP: half wave plate, PBS: polarized beam splitter, SLM: spatial light modulator, Obj.: microscope objective, SM: scattering medium, 2PF: 2-photon fluorescence, BS: beam splitter, M: mirror, L: lens, DM: dichroic mirror, F: filter, P: polarizer.}
\label{fig_setup}
\end{figure}

\begin{figure}[t!]
\centering
\includegraphics[width=\linewidth]{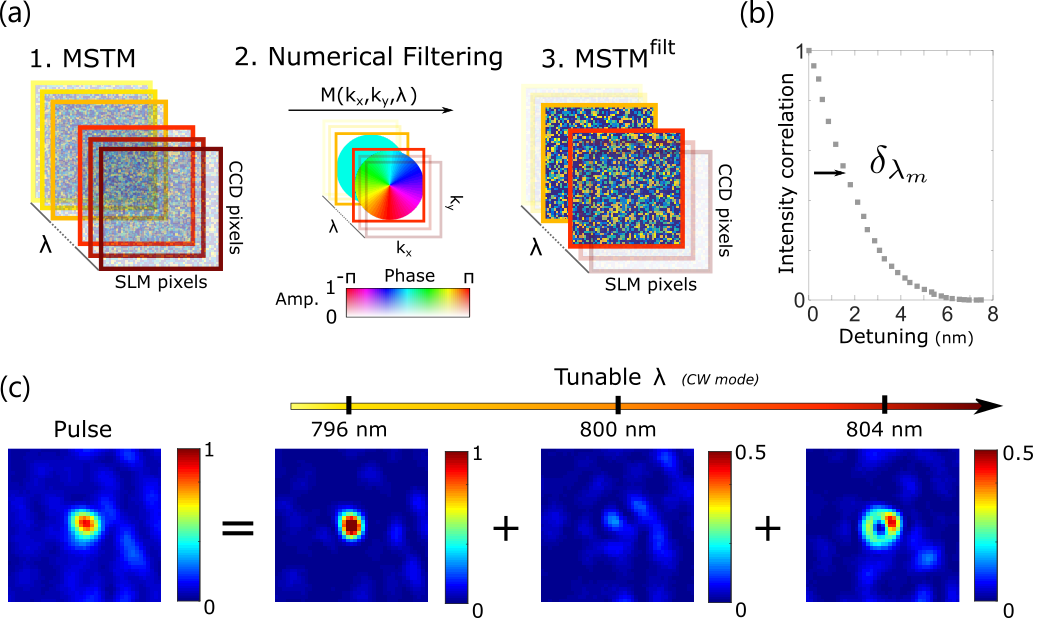}
\caption{Principles of spectrally-dependent PSF engineering. (a) Fourier filtering of MSTM with a spectrally-dependent mask $M(k_x,k_y,\lambda)$ gives rise to a new operator, denoted MSTM$^{\text{filt}}$, which can be used as standard MSTMs. (b) Spectral correlation bandwidth of the scattering medium. Its half width at half maximum (HWHM) is $\delta{\lambda_m}\simeq 1.6$~nm. (c) Experimental focusing at two wavelengths with two different and arbitrary chosen PSFs.}
\label{fig_MS}
\end{figure}

\begin{figure}[t!]
\centering
\includegraphics[width=\linewidth]{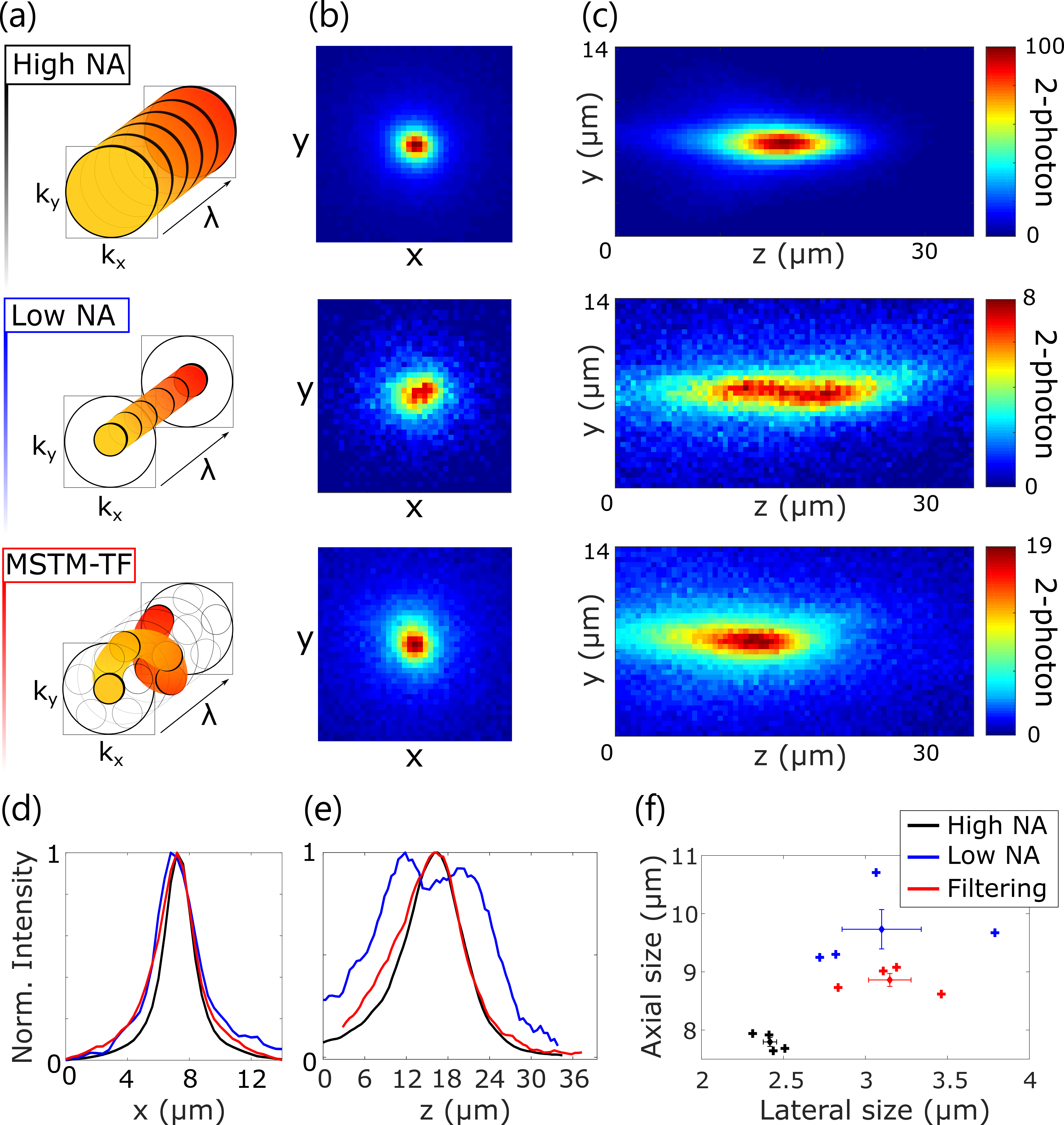}
\caption{Experimental demonstration of temporally focused light generation with a thick scattering medium (MSTM-TF). It combines two PSFs, termed High NA and Low NA (a) Spectrally-dependent masks computed onto the virtual pupil accessible from the Fourier transform of the output field, captured in the MSTM. (b) Lateral profiles of the corresponding generated PSFs (projection of the beam onto z-axis). (c) Axial profiles (projection of the beam onto y-axis).
(d) Lateral profiles (projection of (b) onto y-axis). (e) Axial profiles (projection of (c) onto x-axis). (f) Lateral and axial sizes of each PSF for 4 different spatial positions. We also report the mean and standard deviation values for each case.}
\label{ResultsTF}
\end{figure}

\begin{figure}[t!]
\centering
\includegraphics[width=\linewidth]{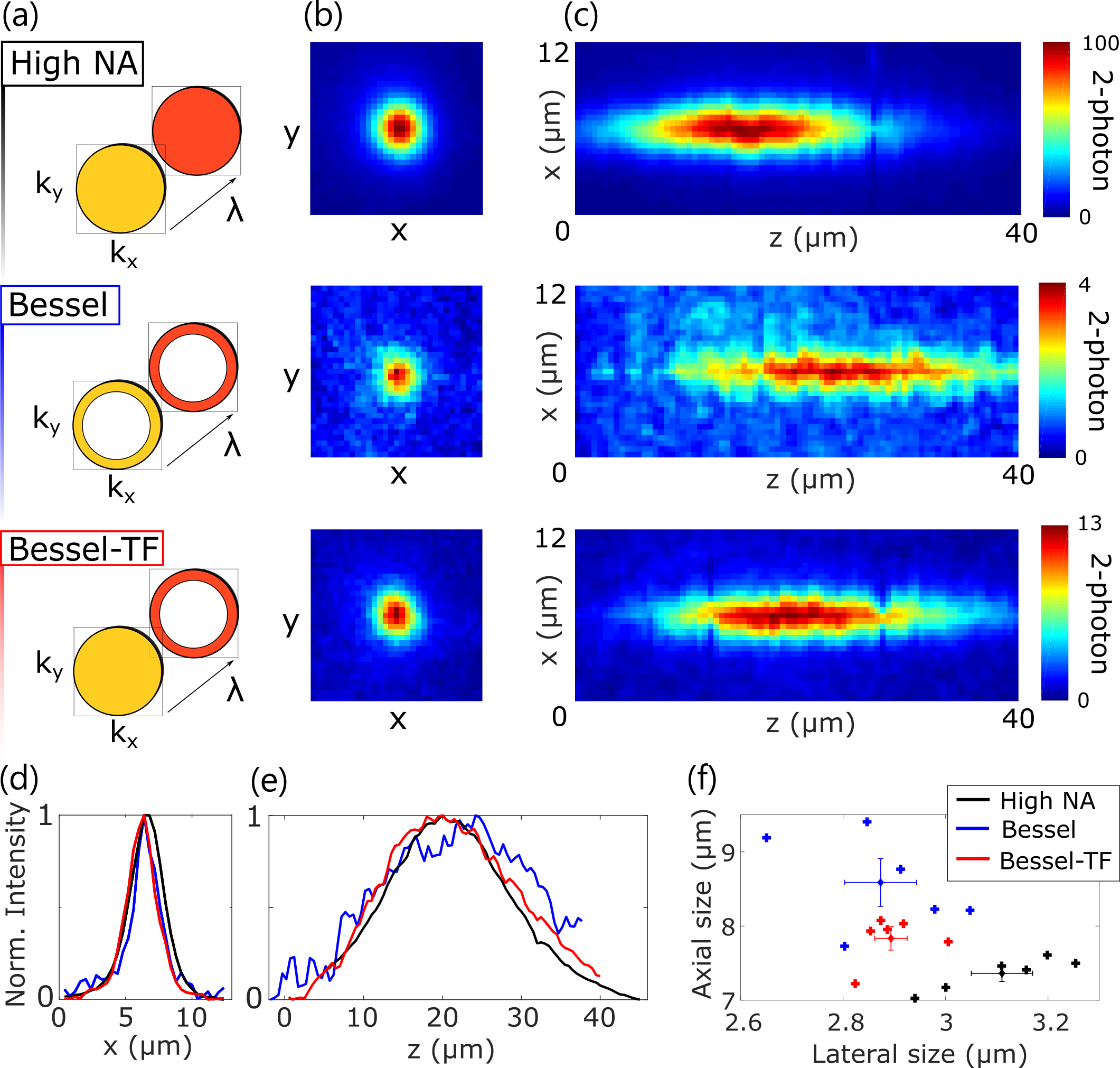}
\caption{Another example of temporally-focused beam with a scattering medium (Bessel-TF). It combines two PSFs, denoted High NA and Bessel -- Experimental Results -- (a) Spectrally-dependent masks computed onto the virtual pupil accessible from the Fourier transform of the output field, captured in the MSTM. The annulus ring has an inner radius $= 0.67 k_1 $, where $k_1$ is the outer radius (i.e. size of the aperture). (b) Lateral profiles of the corresponding generated PSFs (projection of the beam onto z-axis). (c) Axial profiles (projection of the beam onto y-axis). (d) Lateral profiles (projection of (b) onto y-axis). (e) Axial profiles (projection of (b) onto x-axis). (f) Lateral and axial sizes of each PSF for 6 different output spatial positions. We also report the mean and standard deviation values for each case.}
\label{ResultsBes}
\end{figure}

A schematic representation of the experimental optical system is shown in Fig. \ref{fig_setup}. It allows to measure the MSTM, generate the multi-spectral PSF and characterize it in three dimensions. A Ti:Sapphire laser source (MaiTai, Spectra Physics, $120$~fs pulselength, $800$~nm center wavelength,  $12.7$~nm spectral bandwidth) is used either as a tunable monochromatic source, for characterization, or as a pulsed source with its full bandwidth. Here, the MSTM is measured by sweeping the wavelength as in Ref. \cite{Andreoli2015}, but can also be performed using hyper-spectral imaging \cite{boniface2019rapid}. The beam is split into two distinct paths: a reference path and a probe path. In the probe path, a phase-only SLM ($512 \times 512$ pixels, Meadowlark) subdivided in $64\times64$ macropixels is conjugated with the back focal plane of a microscope objective (Olympus Plan N, 20$\times$, NA 0.40), which illuminates a scattering medium made of ZnO nanoparticles (thickness $\sim 100~\mu$m). The transmitted speckle is collected with another microscope objective (Zeiss EC Plan-Neofluar, 40$\times$, NA~1.3). The probe beam leaving the medium is recombined with the reference on a beam splitter (BS) and the hologram is recorded on a CCD camera (CAM1, Allied Vision, Manta G-046). The MSTM elements are obtained by phase-step interferometry of the probe with the reference arm. As the reference and signal arms do not share a common path, phase drifts and fluctuations between them due to temperature gradients and airflow have to be considered. To account for this, the phase drift was monitored and corrected for as shown previously \cite{ploschner2014gpu}. Once measured, the MSTM was stable for few hours after acquisition. This operator, in conjunction with the SLM, enabled us to focus light both spatially and spectrally. The corresponding spatio-spectral PSF was characterized by looking at the scattered pulse on CAM1 as in Fig. \ref{fig_MS}. To infer on the temporal properties of the PSF (Fig. \ref{ResultsTF} and \ref{ResultsBes}), we used a 2-photon fluorescence process. For this purpose, a solution made of fluorescein is placed right behind the scattering medium. The fluorescent sample, together with the collection objective, are mounted on a translation stage in order to characterize the beam profile along the z-axis. The 2-photon fluorescence is recorded with an EMCCD camera (CAM2, Andor iXon 3), placed after a dichroic mirror to separate the probe beam from the 2-photon fluorescence, together with additional filters to filter out the strong SHG signal emitted by the ZnO medium (longpass, FELH0450, Thorlabs) and to block the probe beam (shortpass, $2 \times$ FESH0650, Thorlabs).

To engineer the multi-spectral PSF of transmitted light, we build a new operator from the experimentally measured MSTM by numerically filtering the pupil (see Fig. \ref{fig_MS}a), as previously shown with monochromatic light \cite{boniface2017transmission}. Briefly, we compute the spatial Fourier transform of the experimentally measured output fields, for all the $N_{SLM}$ input modes and $N_{\lambda}$ spectral components. Then the pupil is engineered with a spectrally dependent mask $M(k_x,k_y,\lambda)$, allowing for multi-spectral PSF engineering through thick scattering media. We must specify here that both phase and/or amplitude of the mask are tunable. Finally, an inverse Fourier transform of the filtered pupil function is performed to return to real space. The latter corresponds physically to the focal plane where the MSTM was measured. This last operation generates a filtered multi-spectral TM, denoted in the following MSTM$^{\text{filt}}$. 

In Fig. \ref{fig_MS} we demonstrate the MSTM$^{\text{filt}}$ capability through a proof-of-concept experiment. The scattering medium used here provides $N_{\lambda} = 8$ spectral degrees of freedom, and thus a set of 8 monochromatic TMs are measured to describe the pulse propagation through the medium. For the sake of demonstration, we filtered only two TMs over the $N_{\lambda}$ available, respectively with, a flat ($\lambda=796$~nm) and a spiral ($\lambda=804$~nm) phase mask. 
Phase conjugation of the two TMs at $\lambda=796$~nm and $\lambda=804$~nm enables arbitrary spatial focusing of the two wavelengths on either the same spatial position or on two separate positions. Here the two wavelengths are spatially focused on the same position. As detailed in \cite{Mounaix2016} the two input fields calculated from phase conjugation are algebraically summed and the resulting phase is displayed onto the SLM. As shown in Fig. \ref{fig_MS}c, such shaping focuses the transmitted pulse on the camera. A scan in wavelength exhibits a diffraction-limited focus at $\lambda = 796$~nm and a donut-like shape at $\lambda=804$~nm in agreement with the masks applied in k-spaces of the corresponding matrices. For the other spectral components, no specific focus is obtained: at these corresponding wavelengths the SLM hologram generates a speckle pattern. As an example, we report the output intensity at $\lambda=800$~nm. Similar results were also obtained using a metasurface but it offers a much lower spectral resolution \cite{zhao2015multispectral}. One advantage of our approach is its high degree of reconfigurability; with the same medium and by simply changing the incident wavefront with the SLM, a new spatio-spectral PSF can be generated. 
\newline

\begin{figure}[b!]
\centering
\includegraphics[width=\linewidth]{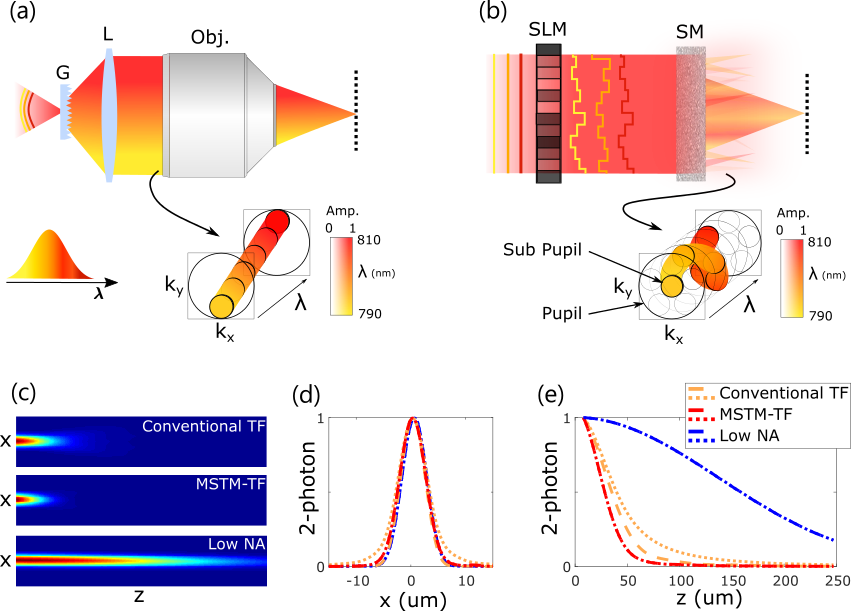}
\caption{
    Temporally focused (TF) light-shaping techniques. (a) Conventional TF. (Top panel) Schematic view of the conventional TF using a grating and an objective lens. (Bottom panel) The pulse spectrum and its spatial unidirectional spreading in the pupil plane (corresponding physically to the back focal plane of the objective). (b) MSTM-TF. (Top panel) Scheme of our MSTM approach based on a scattering medium and a SLM. Each spectral component of the pulse is independently controlled with a SLM. (Bottom panel) The mask applied in a virtual pupil (accessible by computing a Fourier transform). (c) Comparison of the two methods with numerical simulations. While the lateral profile (d) of the two TF methods is similar to the case without shaping (Low NA) their axial confinement (e) is significantly improved. Regarding planes yz (dasked lines) and xz (dotted lines), MSTM-TF has a more uniform PSF than the conventional approach. G: grating, L: lens, Obj.: microscope objective, SLM: spatial light modulator, SM: scattering medium.}
\label{fig_TFprinciple}
\end{figure}
In microscopy, a widespread application of pulse shaping is simultaneous spatial and temporal focusing (TF) \cite{tal2005improved,zhu2005simultaneous}. It consists in shaping the beam in such a way that its pulse length varies along the propagation direction, and that the shortest length occurs only close to the focal plane of the illumination objective (Fig. \ref{fig_TFprinciple}a). It improves significantly the axial sectioning of the 2-photon fluorescence excitation, with various applications in neuroscience \cite{papagiakoumou2008patterned,schrodel2013brain}, since the 2-photon signal is inversely proportional to the pulse length. The key optical part in temporal focusing is its dispersive element which generates the spatio-spectral coupling. Generally, a diffraction grating is used as dispersive element \cite{papagiakoumou2020scanless} but a digital micromirror device has also been used \cite{da2018fast}. In both cases, the spatial dispersion is done along a single axis (see Fig. \ref{fig_TFprinciple}a), with an objective lens focusing all spectral components at its focal plane, where all wavelengths overlap and the shortest pulse length is reached. As such, the technique only applies in free space where the spatio-spectral coupling induced by the grating is known.

In our method, presented in Fig.~\ref{fig_TFprinciple}b, the diffraction grating is replaced by a thick scattering medium which naturally provides the spatio-spectral coupling. Its control is then achieved with a SLM and knowledge of the MSTM.
Once the matrix is measured, the pupil function is numerically filtered with a spectrally-dependent mask $M(k_x,k_y,\lambda)$. The mask is obtained by dividing the pupil (corresponding to the entire accessible k-space) into $N_{\lambda}$ sub pupils,  each defining also a spectral component of the pulse. For instance, the mask $M$ may have an helical shape as represented in Fig. \ref{fig_TFprinciple}b. Importantly, this scheme exploits the entire two-dimensional k-space compared to the conventional unidirectional mapping. It translates into a more isotropic PSF along the z-axis as reported on Fig. \ref{fig_TFprinciple}e. \newline 

To show the versatility of MSTM$^{\text{filt}}$ we report on the experimental realization of two different temporally focused beams which both aim at decoupling axial and lateral resolution. In a first example, we implement the MSTM-TF introduced in Fig. \ref{fig_TFprinciple}b that improves the axial confinement. In a second example, we reduce the axial extension of a Bessel-like beam with a different pupil mask.

As a first experimental demonstration, we generate TF using thick scattering media. As explained in Fig. \ref{fig_TFprinciple}b, our approach consists in filling the pupil, or entire k-space, with $N_{\lambda}$ sub-pupils. The idea is to spectrally combine the spatial properties of two PSFs, coined High NA and Low NA, into a third one, MSTM-TF, which aims at achieving temporal focusing. More specifically, these three PSFs are obtained from the following pupil functions:

\renewcommand{\labelitemi}{$\triangleright$}
\begin{itemize}
    \item High NA, the pupil mask fills the whole aperture (see top panel Fig. \ref{ResultsTF}a). This mask is applied on all the TMs, regardless their wavelength. This mask produces a beam with waist $w_0$ and Rayleigh length $z_R$. The two are related through the numerical aperture $\text{NA} \propto w_0/z_R$.
    
    \item Low NA, the same filtering is done with sub-pupil, three times smaller than the pupil (see middle panel Fig. \ref{ResultsTF}a). Compared to the previous situation this mask produces a beam with larger waist and Rayleigh length. 
    
    \item MSTM-TF, sub-pupils (same sizes as Low NA ones) are centered at a different position for different wavelengths, inside the pupil. Sub-pupils are positioned in such a way that they cover the pupil (see bottom panel Fig. \ref{ResultsTF}a). This mask produces a beam with a waist comparable to Low NA but a reduced Rayleigh length.   
\end{itemize}

The three cases are experimentally compared in Fig. \ref{ResultsTF}. The scattering medium is characterized by $\delta{\lambda_m}\simeq 1.8$~nm. Consequently, for our probe spectrum we have $N_{\lambda}=6$, hence a set of 6 monochromatic TMs were measured to control the full propagation of the pulse spectrum through the medium.
Comparison of lateral (projection onto the z-axis) and axial profiles (projection onto the y-axis) in each case highlights the desired final effect. Quantitatively, we estimate lateral and axial sizes with the 2$^{nd}$ order cumulant of the corresponding distribution profiles. With the same MSTM we repeat the procedure and focus light on three other output spatial positions. Results are plotted on Fig. \ref{ResultsTF}f demonstrating the interest of using such pupil mask function: whereas the lateral size of the MSTM-TF filtering is very similar to Low NA, its axial confinement is improved.
The 2-photon signal through thick scattering media is very weak, leading to long exposure time and limiting the number of positions one can measure within the medium stability time. The full tridimensional characterization of a single PSF (which corresponds to a single point in the graph in Fig. \ref{ResultsTF}f) took approximately $\sim 40$min.

In a second experiment, we combine two other PSFs coined High NA and Bessel, and present another example, Bessel-TF,  for decoupling the lateral and axial profiles of the beam. These three PSFs are obtained from the following pupil functions:

\renewcommand{\labelitemi}{$\triangleright$}
\begin{itemize}
    \item High NA, a large pupil mask is used to filter all the TMs (see top panel Fig. \ref{ResultsBes}a). 
    \item Bessel, the mask corresponds to an annulus with controllable inner and outer radius (see middle panel Fig. \ref{ResultsBes}a). This mask generates a Bessel-like beam whose central lobe is narrower than High NA beam. But this is at the cost of the axial confinement.
    \item Bessel-TF, one TM is filtered with a High NA mask and the other one with a Bessel mask (see bottom panel Fig. \ref{ResultsBes}a). Such pupil creates a tight focus spot with improved confinement compare with the Bessel case.
\end{itemize}

The three cases are experimentally compared in Fig. \ref{ResultsBes}. Their lateral and longitudinal extensions are retrieved with the same post-processing method. As one can notice, the PSF Bessel (middle panel) is very noisy. The associated numerical filtering only retains the high spatial frequencies: all $k > 0.67 k_1$, where $k_1$ is the radius of the pupil. Most of the light (at low spatial frequencies) is rejected and the resulting PSF has a very low signal-to-background ratio (this has been further studied in the Supplementary Material of \cite{boniface2017transmission}). In theory, such PSF is diffraction free, meaning that the longitudinal extent is very large. Due to the background speckle, this expected property is really depreciated. Another downside is that we cannot accurately estimate its lateral and axial sizes (blue crosses in Fig. \ref{ResultsBes}f). However, while a somewhat weaker effect is found for this second example of temporally-focused light shaping, it appears clearly that the spectral combination of High NA and Bessel PSFs provides independent controllable lateral and longitudinal properties.
Since the generation of these PSFs requires only two independent spectral components, we opted for a thinner medium than the one used in the previous experiment, with a $\delta{\lambda_m}\simeq 7$~nm, corresponding to $N_{\lambda}=2$,  and controlled with a set of 2 monochromatic TMs. For the PSF Bessel-TF, the two wavelengths are not equally filtered in terms of energy: only the highest spatial frequencies are used at one $\lambda$ whereas the full pupil is taken at the other one. To compensate for this, we weighted the sum of the two input fields at each wavelength to ensure equal intensity at the focus, which is an additional degree of control allowed by our technique.
Since the medium is thinner, more light is transmitted, which allows speeding up the acquisition. 

Several conditions should be met to successfully and efficiently engineer spatio-temporal PSFs through scattering media. A first important point is the spatio-spectral coupling induced by the scattering material. Our ability to tune the propagation of the beam strongly relies on the number of degrees of freedom, both spatial and spectral, in the system. On the one hand, the number of controlled spatial modes with the SLM translates directly into the quality of the generated beam in the spatial domain. Here we measured the MSTM for a basis of $N_{SLM} = 4096$ orthogonal modes (Hadamard basis), which corresponds to an acquisition time of $\sim 2$~min for a single TM at a given $\lambda$. To retrieve the full MSTM, this operation must be repeated $N_{\lambda}$ times. This is not a major issue here since the medium proved to be stable for few hours. Note that the measurement can be sped up considerably using hyperspectral techniques \cite{boniface2019rapid} or swept-wavelength interferometry \cite{mounaix2019control}. On the other hand, the spectral bandwidth of the medium (and consequently the number of spectral channels $N_{\lambda}$) is a property of the scattering medium itself. $N_{\lambda}$ scales as $\propto (\Delta\lambda^{laser}L^2)/(\lambda^2l_t)$ and can be adjusted with the thickness of the medium $L$ or the transport mean free path $l_t$ \cite{shapiro1986large,aulbachthesis}. However, these quantities also affect  the total transmission of the light $T\propto l_t/L$ in the diffusive regime. In Fig. \ref{ResultsTF} and \ref{ResultsBes}, we characterized the spatio-spectral PSFs by measuring the 2-photon fluorescence signal, which scales as the square of the excitation intensity. Simply put, doubling the thickness of the medium increases $N_{\lambda}$ by a factor 4 but on the other hand the total 2-photon signal is 4 times lower. Therefore in thick scattering media, such as our layers of white paint, there is a trade-off between the transmitted intensity and the number of independent spectral components one can control, which limits the technique to the generation of relatively simple spectral PSFs. However, our approach is very general and may apply to other complex media, such as multimode fibers (MMFs). In a MMF, interference among the guided modes creates wavelength-dependent speckle patterns upon illumination with a coherent source. The spectral correlation width of the speckle $\delta (\lambda)$ scales inversely with the length of the fiber for a fixed numerical aperture \cite{rawson1980frequency,freude1986speckle}, with almost no penalty on the transmitted intensity. Since optical fibers have been optimized for long distance transmission with minimal loss, long fibers can be used to provide very small $\delta (\lambda)$ without altering the total transmission. Such property has already been extensively exploited for turning fibers into high resolution spectrometer \cite{redding2012using,redding2014high} but may be amenable to spatio-spectral pulse shaping with similar spectral resolution.

\vspace{4mm}

In conclusion, we have reported on the formulation of an operator, built upon the experimentally measured MSTM, that enables deterministic spatio-spectral focusing of any arbitrary PSF after propagation through a multiple scattering sample. The spectral resolution is given by the dispersion of the medium and the focusing efficiency by the number of controlled pixels on a single input SLM. We have illustrated the strength of this
technique by characterizing their transverse and longitudinal properties in a temporal focusing application.  The method we propose can readily be extended to other complex media, in particular to multimode fibers that have much higher transmission with similar spectral properties. The possibility of arbitrarily generating complex multi-spectral PSF through multiply scattering media opens up new opportunities in several fields, in particular for microscopy as well as coherent control and nanophotonics.

\section*{Funding Information}
This research has been funded by the European Research Council ERC Consolidator Grant (Grant SMARTIES - 724473). SG is a member of the
Institut Universitaire de France.
HBdA was supported by the LabEx ENS-ICFP: ANR-10-LABX-0010/ANR-10-IDEX-0001-02 PSL*



\newpage
\renewcommand{\refname}{References} 
\bibliographystyle{ieeetr}
\bibliography{BIBLIO}

\begin{thebibliography}{10}

\bibitem{weiner2000femtosecond}
A.~M. Weiner, ``Femtosecond pulse shaping using spatial light modulators,''
  {\em Review of scientific instruments}, vol.~71, no.~5, pp.~1929--1960, 2000.

\bibitem{weiner1990femtosecond}
A.~Weiner, D.~Leaird, G.~P. Wiederrecht, and K.~A. Nelson, ``Femtosecond pulse
  sequences used for optical manipulation of molecular motion,'' {\em Science},
  vol.~247, no.~4948, pp.~1317--1319, 1990.

\bibitem{baxter2006highly}
G.~Baxter, S.~Frisken, D.~Abakoumov, H.~Zhou, I.~Clarke, A.~Bartos, and
  S.~Poole, ``Highly programmable wavelength selective switch based on liquid
  crystal on silicon switching elements,'' in {\em 2006 Optical Fiber
  Communication Conference and the National Fiber Optic Engineers Conference},
  pp.~3--pp, IEEE, 2006.

\bibitem{silberberg2009quantum}
Y.~Silberberg, ``Quantum coherent control for nonlinear spectroscopy and
  microscopy,'' {\em Annual review of physical chemistry}, vol.~60,
  pp.~277--292, 2009.

\bibitem{sun2018four}
B.~Sun, P.~S. Salter, C.~Roider, A.~Jesacher, J.~Strauss, J.~Heberle,
  M.~Schmidt, and M.~J. Booth, ``Four-dimensional light shaping: manipulating
  ultrafast spatiotemporal foci in space and time,'' {\em Light: Science \&
  Applications}, vol.~7, no.~1, pp.~17117--17117, 2018.

\bibitem{PhysRevLett.108.113904}
H.~Vincenti and F.~Qu\'er\'e, ``Attosecond lighthouses: How to use
  spatiotemporally coupled light fields to generate isolated attosecond
  pulses,'' {\em Phys. Rev. Lett.}, vol.~108, p.~113904, Mar 2012.

\bibitem{pariente2015spatio}
G.~Pariente and F.~Qu{\'e}r{\'e}, ``Spatio-temporal light springs: extended
  encoding of orbital angular momentum in ultrashort pulses,'' {\em Optics
  letters}, vol.~40, no.~9, pp.~2037--2040, 2015.

\bibitem{sainte2017controlling}
A.~Sainte-Marie, O.~Gobert, and F.~Quere, ``Controlling the velocity of
  ultrashort light pulses in vacuum through spatio-temporal couplings,'' {\em
  Optica}, vol.~4, no.~10, pp.~1298--1304, 2017.

\bibitem{PhysRevLett.124.134802}
J.~P. Palastro, J.~L. Shaw, P.~Franke, D.~Ramsey, T.~T. Simpson, and D.~H.
  Froula, ``Dephasingless laser wakefield acceleration,'' {\em Phys. Rev.
  Lett.}, vol.~124, p.~134802, Mar 2020.

\bibitem{Zhu:05}
G.~Zhu, J.~van Howe, M.~Durst, W.~Zipfel, and C.~Xu, ``Simultaneous spatial and
  temporal focusing of femtosecond pulses,'' {\em Opt. Express}, vol.~13,
  pp.~2153--2159, Mar 2005.

\bibitem{Oron:05}
D.~Oron and Y.~Silberberg, ``Spatiotemporal coherent control using shaped,
  temporally focused pulses,'' {\em Opt. Express}, vol.~13, pp.~9903--9908, Nov
  2005.

\bibitem{papagiakoumou2020scanless}
E.~Papagiakoumou, E.~Ronzitti, and V.~Emiliani, ``Scanless two-photon
  excitation with temporal focusing,'' {\em Nature Methods}, pp.~1--11, 2020.

\bibitem{Feurer:02}
T.~Feurer, J.~C. Vaughan, R.~M. Koehl, and K.~A. Nelson, ``Multidimensional
  control of femtosecond pulses by use of a programmable liquid-crystal
  matrix,'' {\em Opt. Lett.}, vol.~27, pp.~652--654, Apr 2002.

\bibitem{Vaughan:02}
J.~C. Vaughan, T.~Feurer, and K.~A. Nelson, ``Automated two-dimensional
  femtosecond pulse shaping,'' {\em J. Opt. Soc. Am. B}, vol.~19,
  pp.~2489--2495, Oct 2002.

\bibitem{feurer2003spatiotemporal}
T.~Feurer, J.~C. Vaughan, and K.~A. Nelson, ``Spatiotemporal coherent control
  of lattice vibrational waves,'' {\em Science}, vol.~299, no.~5605,
  pp.~374--377, 2003.

\bibitem{goodman2007speckle}
J.~W. Goodman, {\em Speckle phenomena in optics: theory and applications}.
\newblock Roberts and Company Publishers, 2007.

\bibitem{PhysRevLett.106.103901}
J.~Aulbach, B.~Gjonaj, P.~M. Johnson, A.~P. Mosk, and A.~Lagendijk, ``Control
  of light transmission through opaque scattering media in space and time,''
  {\em Phys. Rev. Lett.}, vol.~106, p.~103901, Mar 2011.

\bibitem{katz2011focusing}
O.~Katz, E.~Small, Y.~Bromberg, and Y.~Silberberg, ``Focusing and compression
  of ultrashort pulses through scattering media,'' {\em Nature photonics},
  vol.~5, no.~6, p.~372, 2011.

\bibitem{morales2015delivery}
E.~E. Morales-Delgado, S.~Farahi, I.~N. Papadopoulos, D.~Psaltis, and C.~Moser,
  ``Delivery of focused short pulses through a multimode fiber,'' {\em Optics
  express}, vol.~23, no.~7, pp.~9109--9120, 2015.

\bibitem{mccabe2011spatio}
D.~J. McCabe, A.~Tajalli, D.~R. Austin, P.~Bondareff, I.~A. Walmsley, S.~Gigan,
  and B.~Chatel, ``Spatio-temporal focusing of an ultrafast pulse through a
  multiply scattering medium,'' {\em Nature communications}, vol.~2, no.~1,
  pp.~1--5, 2011.

\bibitem{Mounaix2016a}
M.~Mounaix, H.~Defienne, and S.~Gigan, ``Deterministic light focusing in space
  and time through multiple scattering media with a time-resolved transmission
  matrix approach,'' {\em Phys. Rev. A}, vol.~94, p.~041802, Oct 2016.

\bibitem{Andreoli2015}
D.~Andreoli, G.~Volpe, S.~Popoff, O.~Katz, S.~Gr{\'{e}}sillon, and S.~Gigan,
  ``{Deterministic control of broadband light through a multiply scattering
  medium via the multispectral transmission matrix},'' {\em Scientific
  Reports}, vol.~5, no.~April, p.~10347, 2015.

\bibitem{popoff2010measuring}
S.~Popoff, G.~Lerosey, R.~Carminati, M.~Fink, A.~Boccara, and S.~Gigan,
  ``Measuring the transmission matrix in optics: an approach to the study and
  control of light propagation in disordered media,'' {\em Physical review
  letters}, vol.~104, no.~10, p.~100601, 2010.

\bibitem{Mounaix2016}
M.~Mounaix, D.~Andreoli, H.~Defienne, G.~Volpe, O.~Katz, S.~Gr\'esillon, and
  S.~Gigan, ``Spatiotemporal coherent control of light through a multiple
  scattering medium with the multispectral transmission matrix,'' {\em Phys.
  Rev. Lett.}, vol.~116, p.~253901, Jun 2016.

\bibitem{Mounaix:18}
M.~Mounaix, D.~M. Ta, and S.~Gigan, ``Transmission matrix approaches for
  nonlinear fluorescence excitation through multiple scattering media,'' {\em
  Opt. Lett.}, vol.~43, pp.~2831--2834, Jun 2018.

\bibitem{boniface2019rapid}
A.~Boniface, I.~Gusachenko, K.~Dholakia, and S.~Gigan, ``Rapid broadband
  characterization of scattering medium using hyperspectral imaging,'' {\em
  Optica}, vol.~6, no.~3, pp.~274--279, 2019.

\bibitem{ploschner2014gpu}
M.~Pl{\"o}schner, B.~Straka, K.~Dholakia, and T.~{\v{C}}i{\v{z}}m{\'a}r, ``Gpu
  accelerated toolbox for real-time beam-shaping in multimode fibres,'' {\em
  Optics express}, vol.~22, no.~3, pp.~2933--2947, 2014.

\bibitem{boniface2017transmission}
A.~Boniface, M.~Mounaix, B.~Blochet, R.~Piestun, and S.~Gigan,
  ``Transmission-matrix-based point-spread-function engineering through a
  complex medium,'' {\em Optica}, vol.~4, no.~1, pp.~54--59, 2017.

\bibitem{zhao2015multispectral}
Z.~Zhao, M.~Pu, H.~Gao, J.~Jin, X.~Li, X.~Ma, Y.~Wang, P.~Gao, and X.~Luo,
  ``Multispectral optical metasurfaces enabled by achromatic phase
  transition,'' {\em Scientific Reports}, vol.~5, no.~1, pp.~1--9, 2015.

\bibitem{tal2005improved}
E.~Tal, D.~Oron, and Y.~Silberberg, ``Improved depth resolution in video-rate
  line-scanning multiphoton microscopy using temporal focusing,'' {\em Optics
  letters}, vol.~30, no.~13, pp.~1686--1688, 2005.

\bibitem{zhu2005simultaneous}
G.~Zhu, J.~Van~Howe, M.~Durst, W.~Zipfel, and C.~Xu, ``Simultaneous spatial and
  temporal focusing of femtosecond pulses,'' {\em Optics express}, vol.~13,
  no.~6, pp.~2153--2159, 2005.

\bibitem{papagiakoumou2008patterned}
E.~Papagiakoumou, V.~De~Sars, D.~Oron, and V.~Emiliani, ``Patterned two-photon
  illumination by spatiotemporal shaping of ultrashort pulses,'' {\em Optics
  Express}, vol.~16, no.~26, pp.~22039--22047, 2008.

\bibitem{schrodel2013brain}
T.~Schr{\"o}del, R.~Prevedel, K.~Aumayr, M.~Zimmer, and A.~Vaziri, ``Brain-wide
  3d imaging of neuronal activity in caenorhabditis elegans with sculpted
  light,'' {\em Nature methods}, vol.~10, no.~10, p.~1013, 2013.

\bibitem{da2018fast}
Y.~Da~Sie, C.-Y. Chang, C.-Y. Lin, N.-S. Chang, P.~J. Campagnola, and S.-J.
  Chen, ``Fast and improved bioimaging via temporal focusing multiphoton
  excitation microscopy with binary digital-micromirror-device holography,''
  {\em Journal of biomedical optics}, vol.~23, no.~11, p.~116502, 2018.

\bibitem{mounaix2019control}
M.~Mounaix and J.~Carpenter, ``Control of the temporal and polarization
  response of a multimode fiber,'' {\em Nature communications}, vol.~10, no.~1,
  pp.~1--8, 2019.

\bibitem{shapiro1986large}
B.~Shapiro, ``Large intensity fluctuations for wave propagation in random
  media,'' {\em Physical review letters}, vol.~57, no.~17, p.~2168, 1986.

\bibitem{aulbachthesis}
J.~Aulbach, {\em Spatiotemporal control of light in turbid media}.
\newblock PhD thesis, University of Twente, 9 2013.

\bibitem{rawson1980frequency}
E.~G. Rawson, J.~W. Goodman, and R.~E. Norton, ``Frequency dependence of modal
  noise in multimode optical fibers,'' {\em JOSA}, vol.~70, no.~8,
  pp.~968--976, 1980.

\bibitem{freude1986speckle}
W.~Freude, C.~Fritzsche, G.~Grau, and L.~Shan-da, ``Speckle interferometry for
  spectral analysis of laser sources and multimode optical waveguides,'' {\em
  Journal of lightwave technology}, vol.~4, no.~1, pp.~64--72, 1986.

\bibitem{redding2012using}
B.~Redding and H.~Cao, ``Using a multimode fiber as a high-resolution, low-loss
  spectrometer,'' {\em Optics letters}, vol.~37, no.~16, pp.~3384--3386, 2012.

\bibitem{redding2014high}
B.~Redding, M.~Alam, M.~Seifert, and H.~Cao, ``High-resolution and broadband
  all-fiber spectrometers,'' {\em Optica}, vol.~1, no.~3, pp.~175--180, 2014.

\end{thebibliography}

\end{document}